\documentclass[%
 reprint,
 amsmath,amssymb,
 aps,
pre,
showkeys
]{revtex4-2}

\usepackage{graphicx}
\usepackage{dcolumn}
\usepackage{bm}

\usepackage{amsmath,amsfonts,amscd,amssymb}
\usepackage{graphicx}
\usepackage{epstopdf}
\usepackage{overpic}
\usepackage{cancel}
\usepackage{rotating}
\usepackage{url}
\usepackage{caption}
\usepackage{color}
\usepackage{rotating}
\usepackage{multirow}
\usepackage{wrapfig}
\usepackage{mathtools}
\usepackage{subeqnarray}
\usepackage{setspace}
\usepackage{hyperref}

\usepackage{natbib}

\graphicspath{{Figures/}}

\begin{document}

\title{Principal component trajectories for modeling spectrally-continuous dynamics as forced linear systems}

\author{Daniel Dylewsky}
 \affiliation{Department of Physics, University of Washington, Seattle, WA 98195.}
 \email{dylewsky@uw.edu}

\author{Eurika Kaiser}
\affiliation{%
 Department of Mechanical Engineering, University of Washington, Seattle, WA 98195
}%

\author{Steven L. Brunton}
\affiliation{%
 Department of Mechanical Engineering, University of Washington, Seattle, WA 98195
}%

\author{J. Nathan Kutz}
\affiliation{%
 Department of Applied Mathematics, University of Washington, Seattle, WA 98195
}%


\begin{abstract}
Delay embeddings of time series data have emerged as a promising coordinate basis for data-driven estimation of the Koopman operator, which seeks a linear representation for observed nonlinear dynamics. Recent work has demonstrated the efficacy of Dynamic Mode Decomposition (DMD) for obtaining finite-dimensional Koopman approximations in delay coordinates. In this paper we demonstrate how nonlinear dynamics with sparse Fourier spectra can be (i) represented by a superposition of {\em principal component trajectories} (PCT) and (ii) modeled by DMD in this coordinate space. For continuous or mixed (discrete and continuous) spectra, DMD can be augmented with an external forcing term. We present a method for learning linear control models in delay coordinates while simultaneously discovering the corresponding exogeneous forcing signal in a fully unsupervised manner. This extends the existing DMD with control (DMDc) algorithm to cases where a control signal is not known {\em a priori}. We provide examples to validate the learned forcing against a known ground truth and illustrate their statistical similarity. Finally we offer a demonstration of this method applied to real-world power grid load data to show its utility for diagnostics and interpretation on systems in which somewhat periodic behavior is strongly forced by unknown and unmeasurable environmental variables.
\end{abstract}

\keywords{model discovery, dynamical systems, dynamic mode decomposition, time delay embedding, Koopman theory, linear control}
\maketitle


\section{Introduction}
Modern time series analysis has been transformed by emerging and innovative mathematical methods from machine learning and data science. The ubiquitous availability of measurement data across the sciences, coupled with advances in storage and processing power, has led to a resurgence of interest in the question of how diagnostic and predictive models can be discovered directly from time series data. Using a dynamical systems perspective allows one to take advantage of a broad variety of existing systems analysis methods with applications to forecasting and control. Of particular interest are techniques which are able to map an observed measurement series into a space where it can be accurately reproduced by a set of linear governing equations. This is the central tenet of Koopman theory, first introduced in 1931, which states that nonlinear dynamical systems can be reproduced by linear evolution in a space of observables on the state variables~\cite{koopman1931,koopmanvonneumann1932}. However, this linear representation typically comes at the price of dimensionality: finite-dimensional nonlinear evolution generically requires an infinite-dimensional lifting to be accurately recreated by a linear operator. Linear models are desirable because they allow for the use of many powerful linear algebraic methods for estimation, prediction, and control, all of which can be extended to nonlinear systems if one is able to construct an accurate finite-dimensional coordinate system for the Koopman operator~\cite{Brunton2019book}. As we show, time-delay embeddings provide such a coordinate system, providing a mathematical framework for the extraction of {\em principal component trajectories} (PCT) which allows for the construction of dynamics, actuated or not, via superposition.

Dynamic Mode Decomposition (DMD) is the leading approach to approximating Koopman operators through regression~\cite{Schmid2010jfm,rowleymezicetal2009,Kutz2016book}. Specifically, it offers a means of efficiently discovering a finite-rank linear approximator to nonlinear dynamics directly from observation data, even in very high dimensions. The simplest implementation of the algorithm regresses an operator which acts directly on the measured state variables, but there is no reason, \textit{a priori}, to expect this space to admit a faithful linear representation. Central to the successful implementation of DMD is the problem of identifying a space of Koopman observables into which the dynamics can be lifted to better suit the assumption of linearity \cite{mezic2004,Mezic2005nd,Budivsic2012chaos,Mezic2013arfm}. A variety of approaches have been taken, including constructing libraries of simple monomial functions evaluated on the data and regressed densely \cite{williamskevrekidisrowley2015} or sparsely \cite{Brunton2016pnas,Brunton2016plosone}, kernel methods such as diffusion mapping \cite{Giannakis2019}, and deep learning of coordinate transformations \cite{Takeishi2017nips,Lusch2018,Mardt2018natcomm,Wehmeyer2018jcp}. The method of interest in this work is the use of time-delay embedding of measurement data, which has been used to great effect in the HAVOK~\cite{bruntonproctorkaiserkutz2017} and the subsequent Hankel DMD~\cite{arbabimezic2017} algorithms. 
DMD on time-delay coordinates offers advantageous properties with respect to Koopman approximation which are universal to any input data \cite{kamb18}. 

Time-delay embedding refers to a coordinate transformation in which a time-localized measurement $x(t)$ is augmented by time-shifted copies of itself $x(t-\tau)$. The use of this technique for data-driven modeling dates back to the seminal Takens embedding theorem, which showed that a chaotic attractor can be reconstructed, up to a diffeomorphism, from the time series of a single measured variable \cite{takens1981}. 
The method has since found purchase in a number of widely-used algorithms, including {\em singular spectrum analysis} (SSA) \cite{broomhead1989}, {\em nonlinear Laplacian spectral analysis} (NLSA) \cite{giannakis2012}, and the {\em eigensystem realization algorithm} (ERA) \cite{juangpappa1985}. Particular attention has been devoted to the use of techniques such as singular value decomposition (SVD) to identify dominant modal content of data represented in delay coordinates \cite{hassani07,kamb18}. 
For a time-delayed scalar measurement series, the principal components obtained by SVD form a temporal basis, with functions similar to those in a Fourier or wavelet basis. 
Indeed, it has been demonstrated that for sufficiently long embedding windows the delay coordinate SVD converges to an estimate of the discrete Fourier decomposition \cite{vautardghil1989}. This property suggests a strong compatibility with the Koopman operator-theoretic approach to systems modeling, in which the eigenvalue spectrum of the desired operator has been shown to relate to the same harmonic averages used to compute the Fourier transform \cite{Mezic2005nd}. This connection has been borne out by recent work on time-delay DMD~\cite{turowleyetal2014,bruntonjohnsonojemannkutz2016}, variations on which form analogous models on their SVD projections \cite{bruntonproctorkaiserkutz2017,arbabimezic2017} or on exact Fourier basis projections \cite{panduraisamy2019}.

The fidelity of a DMD model depends heavily on the Fourier spectral properties of the true dynamics. A finite-dimensional linear operator generates dynamics on a discrete set of frequencies determined by its eigenvalues; if the input data has a continuous spectrum it cannot be fully reproduced by DMD (time-delayed or otherwise). Moreover, the number of discrete spectral peaks that can be captured by a linear model depends on its dimension. DMD represents oscillatory modes by complex conjugate eigenvalue pairs, so a rank-$r$ model will admit at most $r/2$ distinct frequency components. This offers one perspective on the motivation for building models in delay coordinates: a discrete eigenvalue spectrum of arbitrary finite length can be obtained simply by embedding additional shifted copies of the data to increase the dimension of the augmented space. It can be shown that for scalar time series, the minimum number of such embeddings is determined fully by the sparsity of the Fourier spectrum \cite{liebertschuster1988,panduraisamy2019}. The continuous spectrum case, however, presents a much greater challenge for Koopman modeling. While a sufficiently dense point spectrum can approximate a continuous one in the high-dimensional embedding limit \cite{Das2019}, the practical utility of this for numerical methods is limited. It has long been suggested \cite{Mezic2005nd} that a Koopman approach might be taken for the ``almost periodic'' (i.e. spectrally discrete) portion of dynamics and augmented somehow to account for the continuous remainder. Previous attempts at this have used intermittent parametric forcing derived from rank-reduced SVD time series \cite{bruntonproctorkaiserkutz2017} or direct translational manipulation of eigenvalues \cite{Lusch2018} to fill out the spectrum. 

Building on the approach of \cite{bruntonproctorkaiserkutz2017}, this paper makes two principal methodological contributions to the problem of modified Koopman representations for such systems. First, we introduce a means of deriving a novel projectional basis for dynamics using time delay methods inspired by SSA. This "PCT" representation offers properties generically favorable to approximation of the Koopman operator \textit{and} a basis which naturally decomposes a signal into its continuous and discrete spectral components. Second, we propose a technique for Koopman modeling of such mixed-spectrum systems using the DMD with Control (DMDc) algorithm~\cite{proctorbruntonkutz2016}, which fits measurement data to a linear control model using a known control signal. We present a fully unsupervised method to learn a control model and a concomitant forcing signal which together fully reproduce the observed dynamics, with discrete-spectrum behavior modeled by endogenous linear evolution and the continuous-spectrum remainder isolated in the exogenous control. This approach is made possible by the projection of dynamics onto the PCT basis, in which components of the dynamics which are not amenable to low-rank Koopman representation are by default relegated to the projectional residual. 

The remainder of this paper is structured as follows: Sec. \ref{sec:delay_embedding} summarizes the process by which delay coordinate representations are obtained from data. The approach is modeled on that of SSA, but focuses on the less common case of spatiotemporal state delay embedding. We present a novel interpretation for the coordinate bases that this generates and illustrate the information-richness of the resultant modes relative to the highly generic output of the same approach on scalar data. Sec. \ref{sec:dmd} discusses the regression of linear DMD models in the learned delay-coordinate space, using data with discrete and mixed spectra. A procedure is introduced for the extraction of an external forcing signal to account for observed nonperiodic dynamics, with results validating its success on systems with known exogenous forcing. Sec. \ref{sec:dmdc} joins this method with the existing DMDc algorithm, allowing these results to be integrated into a single unified linear control model. Finally, Sec. \ref{sec:power_grid} offers an example of how this technique might be applied to a real-world data set as an informative diagnostic tool.

\section{Full-State Embedding: Methods and Interpretations}
\label{sec:delay_embedding}
In this section we outline a procedure for time-delay analysis of a measurement time series. The steps taken are similar to those of SSA. But where SSA typically entails delay embedding on a scalar signal, we extend this approach to the case of a multivariate vector signal and briefly discuss the interpretation of the principal components that this produces.

\begin{figure*}[t]
\centering
\includegraphics[width=0.95\textwidth]{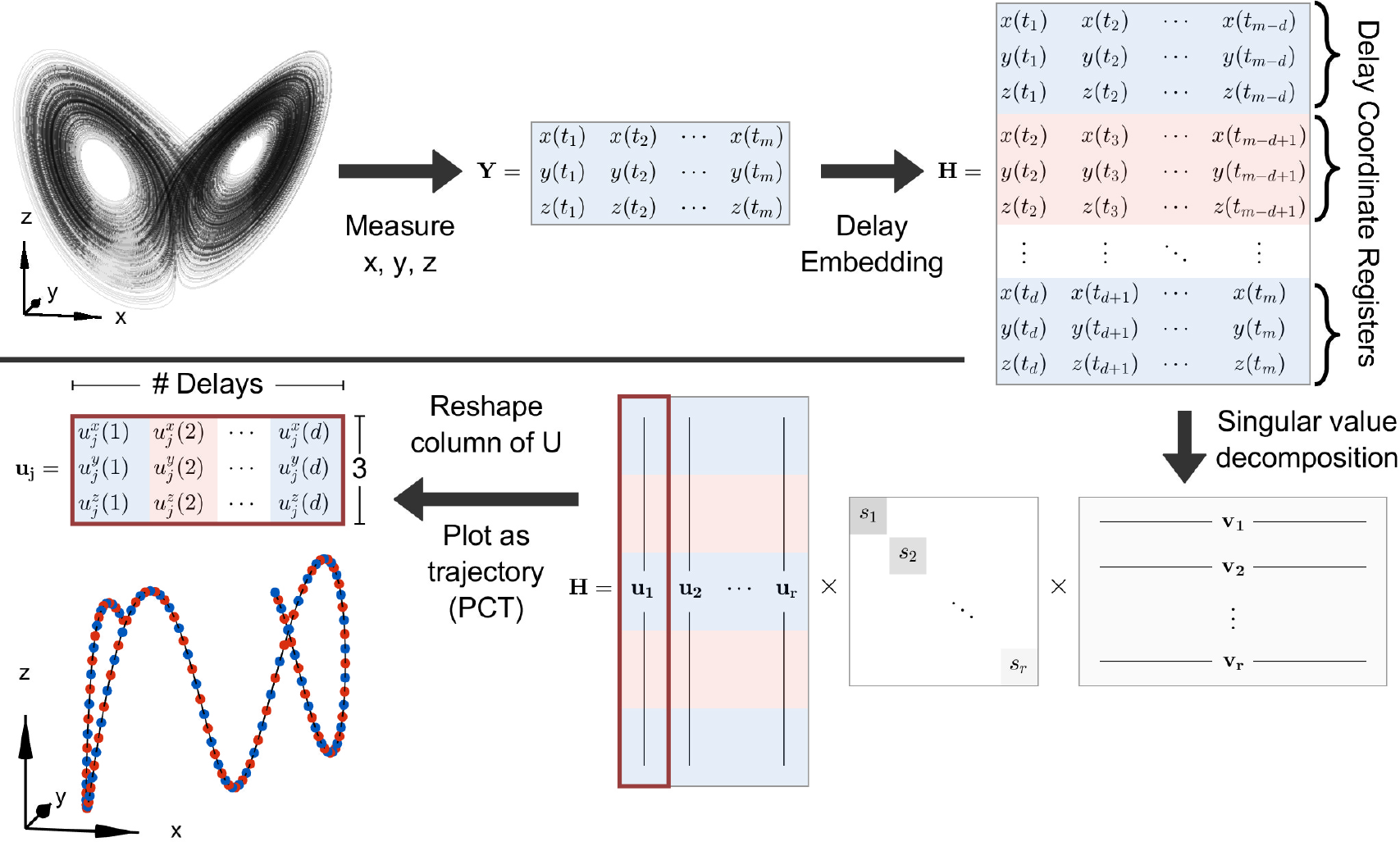}
\caption{Protocol for extracting delay-embedded SVD trajectories, or {\em principal component trajectories} (PCT), from time series data.}
\label{fig:mode_trajectory_protocol}
\end{figure*}

\subsection{Time Delay Embedding}
Delay embedding is the process of lifting a time series signal into a higher dimensional space by stacking it with time-shifted copies of itself. For a scalar $y(t)\in\mathbb{R}$, the lifted Hankel matrix $\mathbf{H}$ is obtained as follows:
\begin{equation}
\begin{split}
\mathbf{Y} &= \left[\begin{matrix} y(t_0) & y(t_1) & \cdots & y(t_m)\end{matrix}\right]
\end{split}
\end{equation}
becomes
\begin{equation}
\begin{split}
\mathbf{H} &= \left[\begin{matrix} y(t_0) & y(t_1) & \cdots & y(t_{m-d}) \\
									 y(t_1) & y(t_2) & \cdots &y(t_{m-d+1})\\
									 \vdots & \vdots & \ddots & \vdots\\
									 y(t_d) & y(t_{d+1}) & \cdots &y(t_m)\end{matrix}\right]
\end{split} .
\label{eq:scalar_delay}
\end{equation}
The matrix
${\bf H}$ contains $d$ shifted copies of the original signal ${\bf Y}$, each offset from the last by one time step. If ${\bf Y}$ had dimension $1\times m$, ${\bf H}$ has dimension $d \times (m-d+1)$. Note that there is considerable redundancy in the elements of ${\bf H}$: all elements belonging to diagonal sets $\{H_{ij}\}$ such that $i+j = \rm(const)$ are equal.

The extension of this procedure to a vector signal $\mathbf{y} \in \mathbb{R}^n$ is straightforward. Each column of ${\bf H}$ now consists of $d$ vectors in $\mathbb{R}^n$ stacked on top of each other with the same time shifting scheme as before:
\begin{equation}
\begin{split}
\mathbf{Y} &= \left[\begin{matrix} | & | &  & | \\
									\mathbf{y}(t_0) & \mathbf{y}(t_1) & \cdots & \mathbf{y}(t_m) \\
									| & | &  & | \end{matrix}\right]
\end{split}
\end{equation}
becomes
\begin{equation}
\begin{split}
\mathbf{H} &= \left[\begin{matrix} | & | &  & | \\
									\mathbf{y}(t_0) & \mathbf{y}(t_1) & \cdots & \mathbf{y}(t_{m-d}) \\
									| & | &  & |\\
									 | & | &  & | \\
									\mathbf{y}(t_1) & \mathbf{y}(t_2) & \cdots & \mathbf{y}(t_{m-d+1}) \\
									| & | &  & |\\
									\vdots & \vdots & \ddots & \vdots \\
									 | & | &  & | \\
									\mathbf{y}(t_d) & \mathbf{y}(t_{d+1}) & \cdots & \mathbf{y}(t_m) \\
									| & | &  & |\end{matrix}\right]
\end{split}.
\label{eq:delay_embed}
\end{equation}
The vectorized Hankel matrix is often used for linear system identification in control applications via ERA~\cite{juangpappa1985}. 
This procedure offers two tunable parameters to be independently chosen. Firstly, the number of delay embeddings $d$ performed on the input signal. This determines the dimension of the delay coordinate space into which the columns of ${\bf y}$ are lifted. Previous work on this topic has sought to quantify the lowest value $d$ can take while still admitting a linear dynamical model which faithfully reproduces observed dynamics \cite{broomheadking1986,Gibson1992phD,panduraisamy2019,liebertschuster1988}. For the purposes of this analysis, however, we assume the availability of input data of arbitrary duration, so there is no need for parsimony in the selection of $d$. While a high value of $d$ does lead to a high-dimensional embedding space, our subsequent model-building efforts will be carried out in a rank-reduced basis spanned by $r$ principal component vectors, or the {\em principal componant trajectories} (PCT) of the time-series. Thus a large number of embeddings $d$ does not carry with it any risk of causing the model regression to be underdetermined.

The second parameter to be selected is the embedding period $T^d$. In Eq.~(\ref{eq:delay_embed}) the time lag introduced in each successive embedding is equal to the sampling resolution of ${\bf y}(t)$, so $T^d = t_d-t_0 = d\Delta t$, assuming the columns of ${\bf Y}$ are sampled at evenly spaced intervals of $\Delta t$. However, the time shift per embedding could just as easily be set to any integer multiple of $\Delta t$, leading to a modified embedding period $T^d = qd\Delta t$, where $q\in \mathbb{Z}^+$. The governing principle for making this choice should be to match the time scale of the dynamics of interest: in the low-$T^d$ limit the stacked state vectors which comprise the columns of ${\bf H}$ will be highly redundant, encoding the passage of a time interval too short for any meaningful dynamical evolution of the system. In the high-$T^d$ limit, the delay coordinate sampling resolution is too coarse to resolve the dynamics; unless ${\bf y}(t)$ is perfectly periodic the delay registers of ${\bf H}$ will become completely uncorrelated. If the system in question contains multiscale dynamics with periods separated by orders of magnitude, multiple sets of embedding parameters may be required. A more extensive discussion of this case with respect to the application of dynamic mode decomposition can be found in \cite{dylewsky2019}. For the purposes of this paper, we restrict our analysis to monoscale dynamics and choose $T^d$ such that it spans a few periods of the lowest-frequency peak of the Fourier spectrum of ${\bf y}(t)$. A more detailed discussion of this parameter selection is available in \cite{Gibson1992phD}.

\subsection{SVD and Full-State Embedding}
Of central importance to the signal processing technique of SSA \cite{broomheadking1986,Fraedrich1986,vautardghil1989}, and to more recent dynamical systems work with time delay coordinates \cite{arbabimezic2017,bruntonproctorkaiserkutz2017,kamb18}, is the application of singular value decomposition (SVD) to the delay-embedded Hankel matrix ${\bf H}$ (as defined in Eq.~(\ref{eq:delay_embed})). In this section we present a brief discussion of the motivation for this step, convergence properties of the results, and interpretations of the modes in delay space specifically for the case of full-state vector embedding.

The SVD of a matrix $\mathbf{Y}$ yields $\mathbf{Y} = \mathbf{U} \mathbf{S} \mathbf{V}^T$. If $\mathbf{Y}^{n\times m}$ is a time series with state dimension $n$ and time dimension $m$, then the columns (or \textit{modes}) of $\mathbf{U}$ form an orthonormal basis for the state space, the columns of $\mathbf{V}$ are time series representing normalized projections of the state onto the $\mathbf{U}$ modes, and the elements of the diagonal matrix $\mathbf{S}$ carry the energetic (correlation) weight of each mode's contribution to the full signal $\mathbf{Y}$. Because modal amplitudes are encoded entirely by $\mathbf{S}$, modes can be hierarchically ranked by relative magnitude. In the examples presented in this paper we assume each coordinate of the input data has mean 0, which can be accomplished for arbitrary $\mathbf{Y}$ by applying the transformation $\mathbf{Y}\to\mathbf{Y}-\bar{\mathbf{Y}}$. Under this condition the SVD is equivalent to Principal Component Analysis (PCA), and the singular values can be interpreted as the variance of the data along the axes of the left singular vectors (the columns of $\mathbf{U}$). Truncating the decomposition to contain only the top $r$ modes yields a rank-reduced representation which produces a least-squares optimal reconstruction of $\mathbf{Y}$. [Note that strictly speaking, centering $\mathbf{Y}$ on the origin does not precisely center $\mathbf{H}$; each row of $\mathbf{H}$ contains a truncated subset of the corresponding row in $\mathbf{Y}$. However, we are concerned with the limit of $m>>d$ in which this discrepancy is negligible, so mean subtraction of either matrix yields results which are equivalent in a practical sense.]

The SVD can be applied to a time series lifted into delay coordinates, as in Eq. \ref{eq:delay_embed}, to obtain a decomposition of the same form:
\begin{equation}\label{HankelSVD}
\begin{split}
\mathbf{H} = \mathbf{U} \mathbf{S} \mathbf{V}^T
\end{split}
\end{equation}
Resultant modes have additional degree of interpretability owing to the temporal structure embedded in $\mathbf{H}$ . Columns of $\mathbf{U}$ are vectors in the input state space, which in this case is the high-dimensional delay space. If $\mathbf{H}$ was constructed from a scalar time series $y(t)$, the SVD modes can be treated as $d$-element time series spanning a duration of $T^d$. If $\mathbf{H}$ were instead the full-state embedding of a vector times series $\mathbf{y}(t) \in \mathbb{R}^n$, its modes can be interpreted as trajectories in $\mathbb{R}^n$, also of duration $T^d$. An intuitive parallel for these decompositions can be found in time-frequency analysis methods such as windowed Fourier and wavelet transforms, both of which have been applied to the DMD method~\cite{kutz2016multiresolution,dylewsky2019}. Thus a vector projection of ${\bf H}$ onto the columns of $\mathbf{U}$ in delay coordinates is analogous to a time-localized projection onto some template function, e.g. a wavelet confined to a length of $T^d$.

This procedure is shown in Fig. \ref{fig:mode_trajectory_protocol}. Note that when the input signal is multivariate, vectors in the delay embedding space have elements spanning both spatial coordinates $x,y,z$ and measurement times $t_j, t_{j+1},...,t_{j+d}$. To treat this as a normed vector space (as SVD does) is to make an implicit assumption about the relative weighting of spatial versus temporal coordinate separations. This assumption is carried in the choice of time spacing between delay embeddings, or equivalently in the choice of embedding period $T^d$. A more concrete resolution could be obtained by explicitly defining a space-time inner product in the Hankel vector space as described in \cite{schmidt2019}, which would also offer a more rigorous basis for interpreting singular value weightings as mode energies of the PCTs.

\subsection{Interpreting SVD Modes in Delay Coordinates}
The result of the decomposition process outlined in Fig. \ref{fig:mode_trajectory_protocol} depends heavily on the structure of ${\bf H}$. By construction, elements of ${\bf H}$ are represented redundantly up to $d$ times (specifically, all $H_{ij}$ where $i+n\times j=(\rm{const})$ are equal). This property holds almost regardless of the dynamical content of the signal in question, and leads to certain guarantees on the SVD modes of ${\bf H}$. In particular, it can be shown that for embeddings of a scalar time series ($n=1$), modes converge to simple functions in the limiting cases of $T^d$. For small $T^d$, the columns of ${\bf U}$ resemble Legendre polynomials, while for large $T^d$ they become sinusoidal \cite{broomheadking1986,vautardghil1989}. (An exception to this convergence property would occur in a purely mixing system with continuous spectrum measured with perfect precision; here all but one singular value would approach 0 in the $d\to\infty$ limit. But in the practical case of large-but-finite $d$ this does not pose any problem.) Indeed, in some circumstances exact instantiations of these functions have been used instead of the data-driven approximations obtained with the SVD \cite{Gibson1992phD,panduraisamy2019}. In this paper we focus on the large-$T^d$ limit, in which delay-coordinate SVD can be thought of as reproducing the discrete Fourier transform on a sparsified frequency basis: the near-sinusoidal ${\bf U}$ modes resemble the projection basis used in the DFT, but instead of a large number of basis functions densely populating frequency space, only the (SVD rank) $r$ most prominently represented frequencies are retained.

\begin{figure}[t]
\centering
\includegraphics[width=\columnwidth]{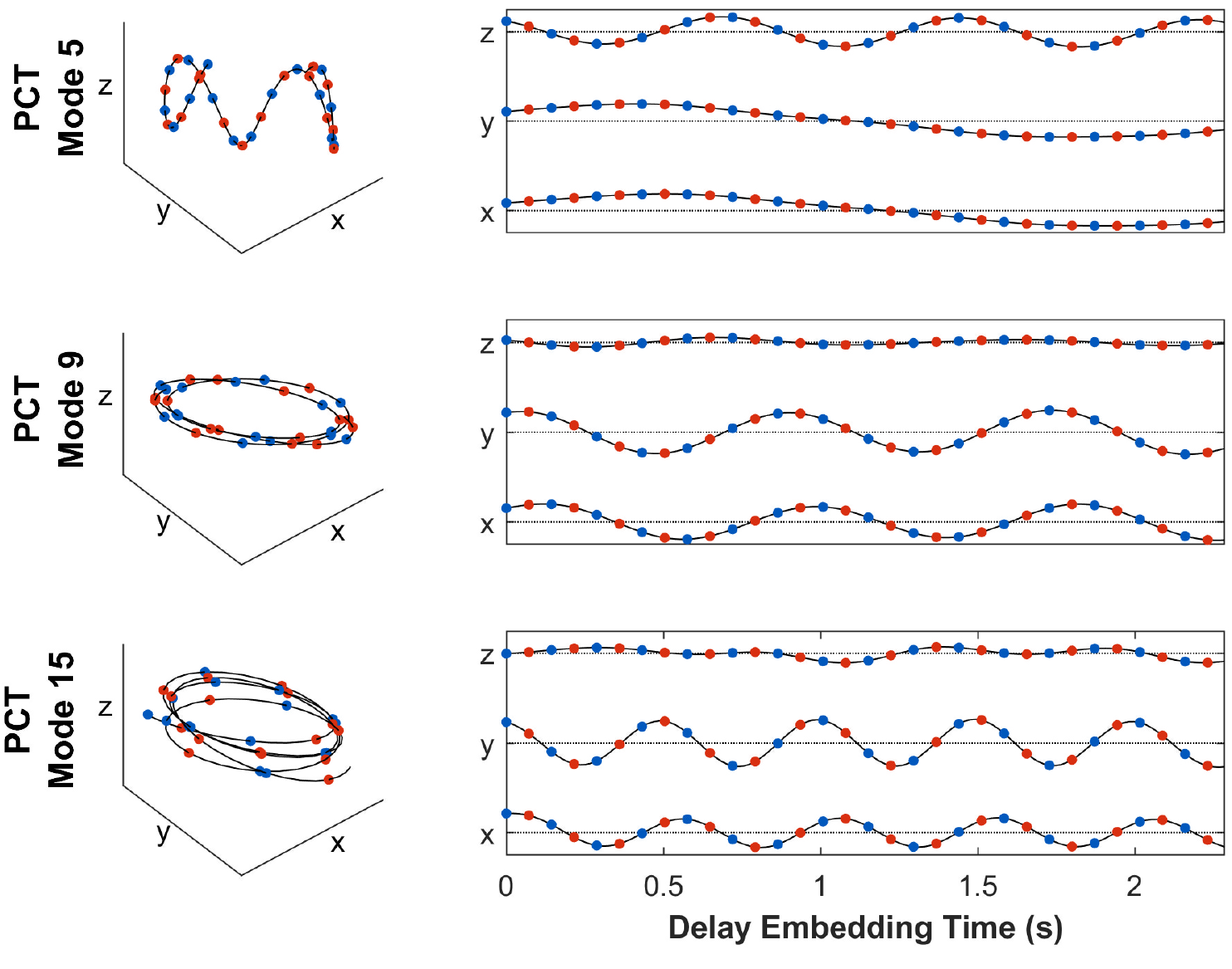}
\caption{Time-delayed SVD modes for a chaotic Lorenz system. Full 3D representations of the PCTs (left) are composed of approximately sinusoidal oscillation in the $x,y,z$ coordinates with (often) incommensurate frequencies and amplitudes.}
\label{fig:mode_trajectory_breakdown}
\end{figure}

Shifting attention to the case of multivariate vector embeddings, we observe similar behavior in the long embedding time limit. Some sample modes obtained for the chaotic Lorenz system are plotted in Fig. \ref{fig:mode_trajectory_breakdown}. The trajectories in $\mathbb{R}^3$ traced out by the modes result from sinusoidal oscillation in $x$, $y$, and $z$. The frequencies of oscillation for different state variables within the same mode are typically equal or related by an integer ratio, but the relative phases and amplitudes are free to vary. Consequently, modes which in the scalar high-$T^d$ case could be completely defined by a single frequency value take on much richer geometric structure in the vector high-$T^d$ case. The figures which can be thus formed by parametric sinusoids belong to the class of generalized $N$-dimensional Lissajous figures \cite{albertazzi2013}, in the special case where frequencies differ by integer ratios. Note that the periods of oscillation are independent of the embedding period $T^d$, so these curves do not in general form closed orbits. 

\begin{figure}[t]
\centering
\includegraphics[width=0.9\columnwidth]{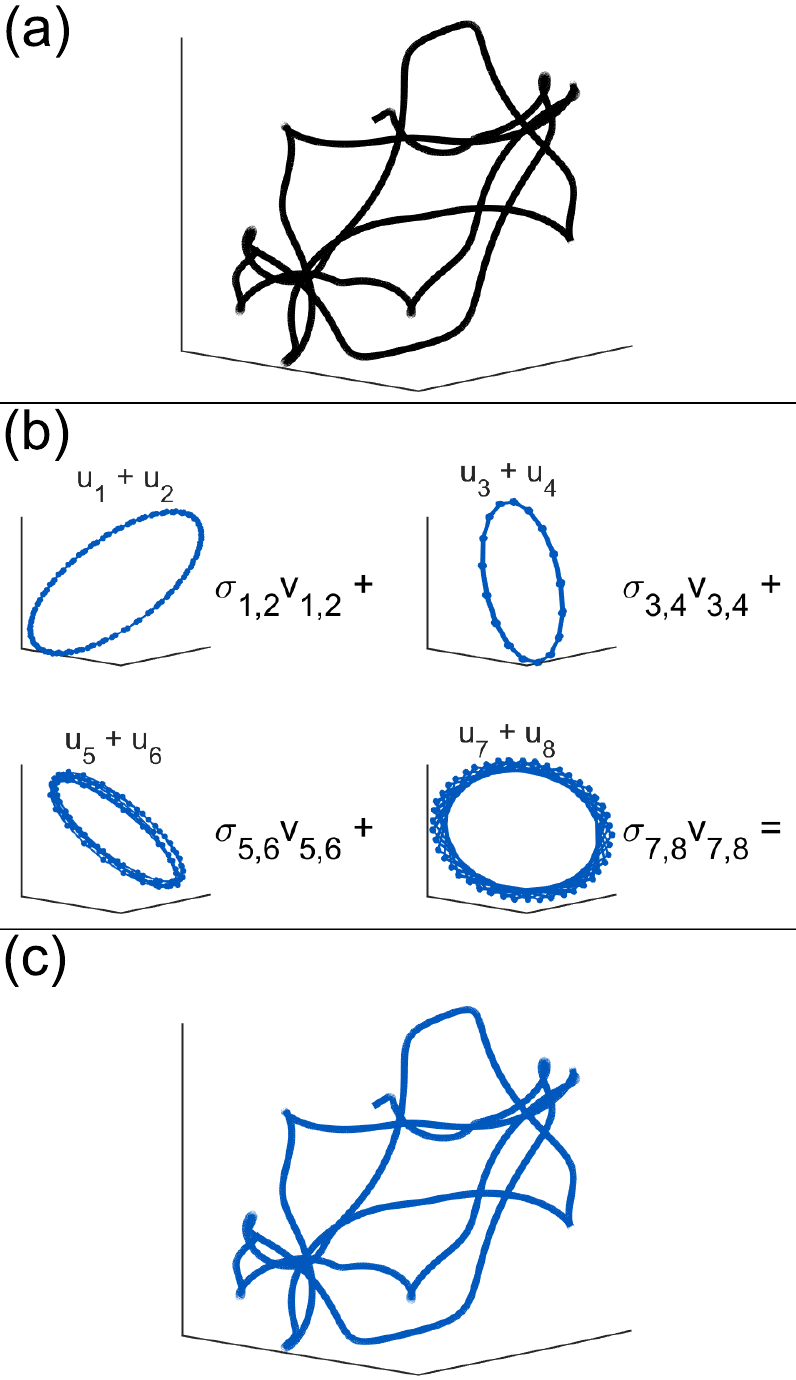}
\caption{Pointwise linear combinations of simple modal orbits are used to reconstruct a complex trajectory in $\mathbb{R}^3$ (a). Modes are grouped together (e.g. $u_1+u_2$) because delay-coordinate SVD naturally identifies pairs of modes which are identical up to a $90^\circ$ phase shift (b). The trajectory reconstituted from these 8 modes is plotted in (c).}
\label{fig:spirograph3D}
\end{figure}

The SVD reconstruction~\eqref{HankelSVD} can be equivalently written as a sum over individual modes:
\begin{equation}
\begin{split}
{\bf H} &= \sum_j {\bf u}_j s_j {\bf v}_j^*
\end{split}
\end{equation}
Each column of ${\bf H}$, which represents a windowed snapshot of the dynamics over a period of $T^d$, is reproduced by a linear combination of the orbits $\{{\bf u}_j\}$ weighted by the singular values $\{s_j\}$ and the time series projections $\{{\bf v}_j(t)\}$. The delay-coordinate SVD therefore functions as an automated means of decomposing a complex trajectory into simple orbital components (Fig. \ref{fig:spirograph3D}), or the PCT coordinates of interest. The result is reminiscent of the Ptolemaic model of the solar system, in which the apparent retrograde motion of planets in the sky is explained using a hierarchy of superimposed epicycles in geocentric coordinates.

It should be noted that although PCTs often look periodic as a result of the Fourier converge property discussed above, they do not correspond to true closed orbits on the original attractor. They should be interpreted as an optimal basis for low-rank projection in the spirit of Principal Component Analysis; the dynamics of the system are unlikely to resemble any given one of the PCTs, but should be reproducible by their linear superposition. The novelty of PCTs relative to traditional PCA and SSA lies in the dynamical richness of the modes. Multivariate time delay embedding lifts data into a high dimensional, hybrid spatial-temporal  representation which admits a modal decomposition which offers a much more compact, low-rank representation for a large class of dynamics than other principal component or Fourier projection methods.

\subsection{PCT of a Nonlinear Oscillator with Discrete Fourier Spectrum}

\begin{figure}[t]
\centering
\includegraphics[width=\columnwidth]{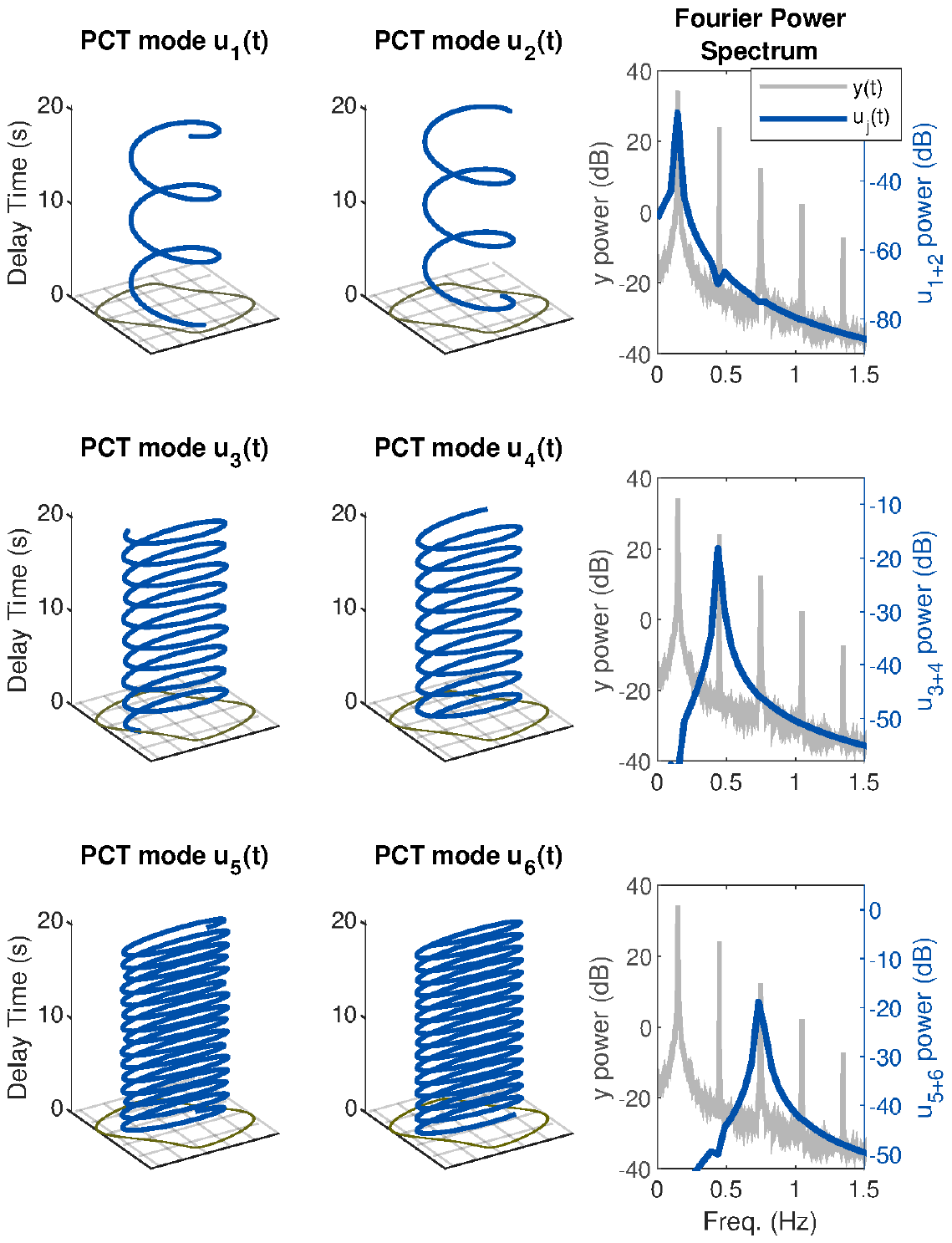}
\caption{First 6 SVD modes for the delay-embedded Van der Pol oscillator. Each mode can be thought of as a trajectory in $\mathbb{R}^2$, with the $z$ coordinate here representing the delay time. They organize into pairs $90^{\circ}$ out of phase in order to form a complete orthogonal basis for a given frequency. Power spectra plotted on the right show that the dominant SVD modes naturally correspond to the frequencies most prominently represented in the signal.}
\label{fig:vdp_svd_modes}
\end{figure}

A more concrete illustration of the analogy to multivariate Fourier decomposition can be found in the example of the Van der Pol system. This canonical nonlinear oscillator is defined by the coupled first-order equations:
\begin{equation}
\begin{split}
\dot{y}_1 &= y_2\\
\dot{y}_2 &= \mu\left(1-y_1^2\right)y_2 - y_1
\end{split}
\label{eq:vdp_def}
\end{equation}
with $\mu$, which controls the strength of the nonlinearity, taken to be $1$ from here on. This system admits solutions which form a closed periodic orbit, so the Fourier spectrum of the trajectory it generates is discrete (only integer harmonics of the base periodic frequency are present). Applying the procedure from Fig. \ref{fig:mode_trajectory_protocol} with $T^d = 20$ produces the delay-coordinate modes plotted in Fig. \ref{fig:vdp_svd_modes}. The state space of Eq. \ref{eq:vdp_def} is represented by the $x-y$ plane, and the time coordinate on the interval $[0, T^d]$ is extended up the $z$ axis to better illustrate frequency distinctions. In the rightmost column, Fourier spectral content of each (combined) mode pair is overlaid on the spectrum of the full Van der Pol trajectory.

The results for this simple example are illustrative of three key features of vector-embedded PCT. First, the modes naturally adhere to the frequencies present in the power spectrum of the input signal. In other words, the PCT not only converge to a sinusoidal basis, but it discovers a sparse Fourier basis made up of those sinusoids most strongly represented in the data. Second, the ordering of the modes corresponds to the heights of their corresponding spectral peaks. This follows from the fact that in the limit as the columns of ${\bf U}$ converge to sinusoids, the singular values associated with them become directly analogous to Fourier power measurements in that both represent a magnitude of the energetic content of the full dynamics contained in a projection onto a given oscillatory mode. Third, the PCT modes naturally self-organize into pairs which share a common frequency and singular value, which is often observed in the fluid flow past bluff bodies~\cite{Noack2003jfm,Taira2017aiaa}. In each row in Fig. \ref{fig:vdp_svd_modes}, the modes are identical up to a $90^\circ$ rotation. This again suggests a parallel to Fourier analysis: a pair of out-of-phase sinusoids of equal frequency (e.g. $\sin(\omega t)$ and $\cos(\omega t)$) form a complete basis for dynamics of arbitrary phase at that frequency. 

\subsection{PCT and Continuous-Spectrum Dynamics}
As the spectral plots in Fig. \ref{fig:vdp_svd_modes} suggest, PCT is a method suited to systems whose dynamics admit a discrete Fourier representation. In the long-embedding limit in which PCT modes converge to sinusoids, a decomposition of finite rank $r$ can at most reproduce $r/2$ spectral peaks in its reconstruction. For a system with a continuous spectrum, this is clearly insufficient: no finite number of discrete frequencies can densely cover a continuous interval on the frequency space. Thus, while the delay method of Fig. \ref{fig:mode_trajectory_protocol} can be applied to any uniformly-sampled time series data in $n$ dimensions, its utility as a means of sparsely representing the dynamical content of the given signal is limited to the discrete-spectrum case.

\begin{figure*}[t]
\centering
\includegraphics[width=\textwidth]{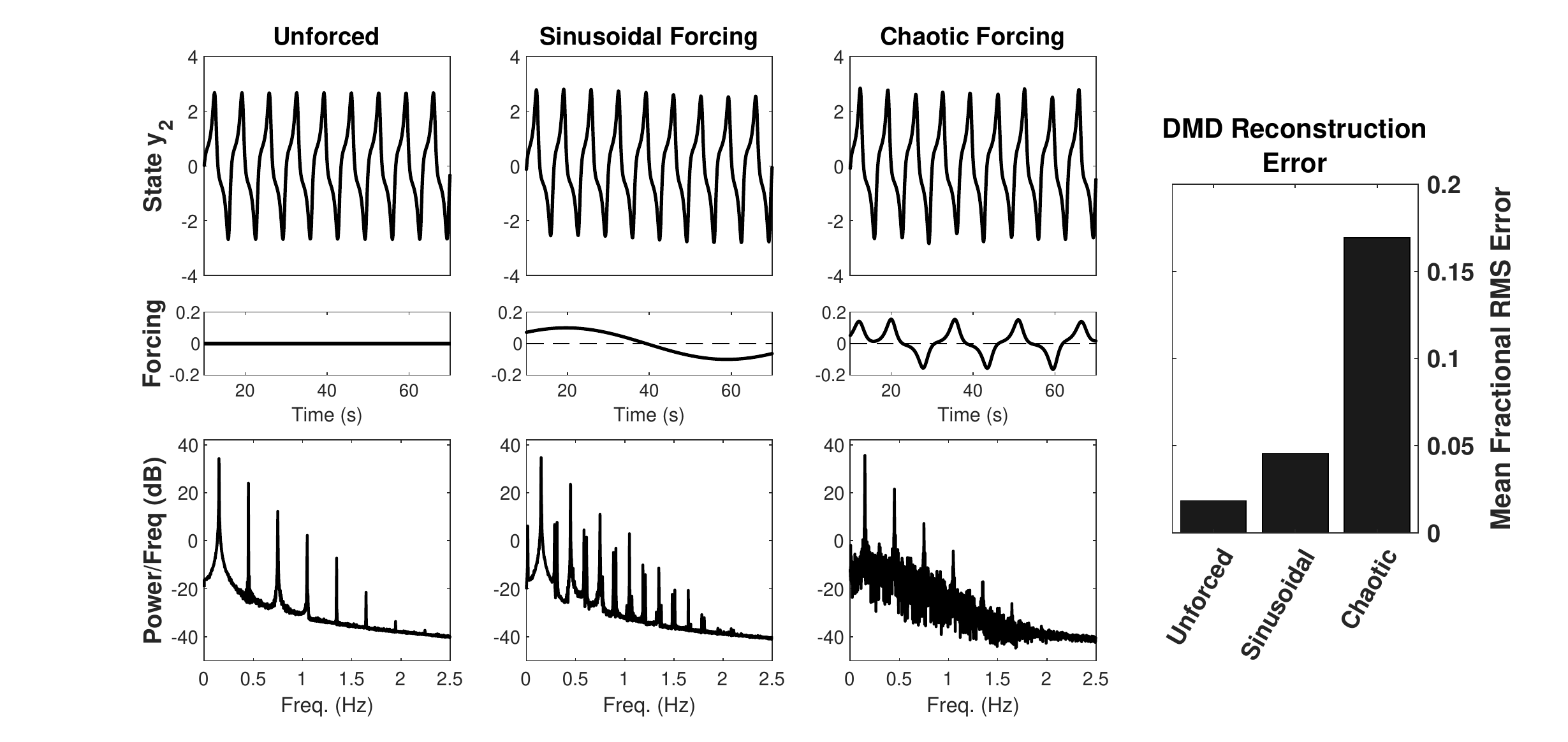}
\caption{Power spectrum of a Van der Pol oscillator system subjected to different types of forcing.}
\label{fig:power_spectrum_comparison}
\end{figure*}

This is a highly restrictive condition, given that many systems of mathematical interest and almost all real-world systems do not exhibit fully discrete Fourier spectra. Even when dynamics are very nearly periodic, the introduction of any measurement noise or stochastic/chaotic forcing will populate the gaps in the spectrum and undermine any attempt to build a linear model for the system in delay coordinates. This is illustrated in Fig. \ref{fig:power_spectrum_comparison}, in which the Van der Pol oscillator from Eq. \ref{eq:vdp_def} is subjected to different types of weak parametric forcing on the $y_2$ variable. Though the magnitude of forcing is sufficiently small that the state-space trajectories (top) look nearly identical, the differences in the power spectra (bottom) are quite obvious. The consequences of this spectral contamination are evident when the plotted time series are subjected to linear model discovery using DMD, which is discussed in the next section. Even when the exogenous forcing is an order of magnitude weaker than the intrinsic dynamics, it yields a model whose reconstruction error is many times greater than that of the unforced system. Because the base periodic oscillation still dominates the observed dynamics, a linear delay-coordinate model should offer a useful approximation to the true behavior. The remainder of this paper is devoted to understanding how such a model can be obtained by isolating the continuous-spectrum forcing from the underlying periodic motion and combining them in the framework of linear control.

\section{Dynamic Mode Decomposition and Time-Delay Embedding}\label{sec:dmd}

\begin{figure*}[t]
\centering
\includegraphics[width=0.9\textwidth]{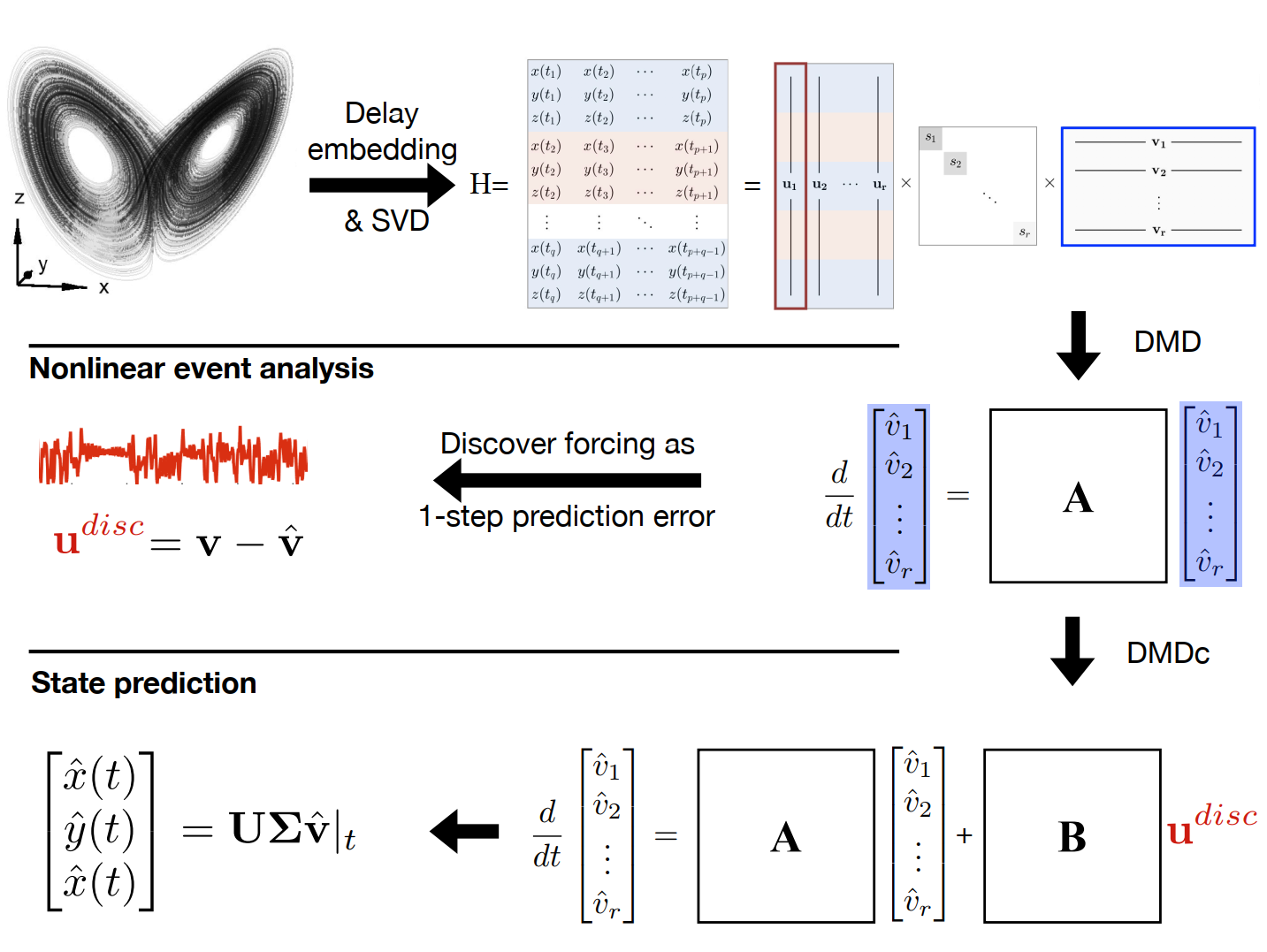}
\caption{Procedure for building linear models from the delay-coordinate SVD illustrated in Fig. \ref{fig:mode_trajectory_protocol}. DMD performed on the time series $V$ matrix yields a linear model whose stepwise prediction error is interpreted as a learned forcing signal $u^{\rm{disc}}$. The same $V$ matrix can then be fed to the DMD with control (DMDc) algorithm along with $u^{\rm{disc}}$ in order to construct a linear control model of the form shown.}
\label{fig:dmd_schematic}
\end{figure*}

\subsection{Dynamic Mode Decomposition}\label{sec:dmd_background}
Dynamic Mode Decomposition (DMD) is a model regression algorithm which produces a best-fit linear operator to reproduce the dynamics observed in some time series data set \cite{Schmid2010jfm,rowleymezicetal2009,Tu2014jcd,Kutz2016book}. The resulting model offers the benefits of interpretability (via its eigenvalue spectrum and spatial eigenvectors) and future-state forecasting to arbitrary time. Given a set of sequential measurements $\mathbf{X} = \left[\mathbf{x}_1, \mathbf{x}_2, ... \mathbf{x}_m\right] \in \mathbb{R}^{n\times m}$, DMD seeks to solve
\begin{equation}
\begin{split}
\dot{\mathbf{X}}= \mathbf{A}\mathbf{X}
\end{split}
\end{equation}
The simplest implementation of the algorithm, known as {\em exact DMD}, simply seeks a least-squares best fit which minimizes $\|\dot{\mathbf{X}} - \mathbf{A}\mathbf{X}\|_F$, solved by $\mathbf{A} = \dot{\mathbf{X}}\mathbf{X}^\dagger$ ($\dagger$ denotes the Moore-Penrose pseudo-inverse) \cite{turowleyetal2014}. This is sometimes formulated in discrete time, where one seeks an operator which maps the state one time step into the future. The resultant $\mathbf{A}$ defines a first-order linear differential equation in the state space $\mathbb{R}^n$, the closed-form solution to which can be written using the operator's eigendecomposition:
\begin{equation}
\begin{split}
\label{eq:dmd_sol}
\tilde{\mathbf{x}}(t) &= \sum_j b_j\mathbf{w}_j e^{\omega_j t}
\end{split}
\end{equation}
where $\mathbf{w}_j$ and $\omega_j$ are eigenvectors and eigenvalues of the DMD operator and $b_j$ are scalar weighting coefficients. This illustrates the power of linear model regression: the obtained DMD model admits a closed-form analytic solution which can easily be broken down into simple exponential modes and can be evaluated over any time domain. DMD has become a heavily used algorithm in the analysis of spatio-temporal data analysis, with many recent innovations including a sparsity promoting variant~\cite{Jovanovic2014pof}, an optimized framework~\cite{askham2018variable}, extensions to control~\cite{proctorbruntonkutz2016}, a Bayesian formulation~\cite{Takeishi2017JCAI}, a variational formulation~\cite{nuske2014jctc,noe2013variational,azencot2019consistent}, a higher-order formulation (HO-DMD)~\cite{leclainchevega2017a}, a stochastic formulation~\cite{Yeung2017arxiv} and a multi-resolution analysis~\cite{kutz2016multiresolution,dylewsky2019}.

\subsection{Time-Delay DMD}
A resurgence of interest in Koopman theory and consequently in DMD in recent years has led to a variety of extensions of the algorithm focusing on two principal tasks: modifying the optimization objective function to redefine what constitutes a ``best fit'' for ${\bf A}$ and identifying lifting transformations which render the data more amenable to linear representation \cite{williamskevrekidisrowley2015,Giannakis2019,Lusch2018}. The results presented in this work remain agnostic on the former issue and focus on a particular solution to the latter: namely, the use of delay-embedding coordinates, or PCT. The lifting procedure described in Eq.~(\ref{eq:scalar_delay}) has been shown to quite generically improve DMD representations for continuous time-series data \cite{Mezic2005nd,leclainchevega2017a}. 
Of particular note is the original Hankel alternative view of Koopman (HAVOK) formulation~\cite{bruntonproctorkaiserkutz2017}, and the subsequent and closely related Hankel DMD~\cite{arbabimezic2017}. 

The procedure for constructing a delay-coordinate linear model is diagrammed in the first two rows of Fig. \ref{fig:dmd_schematic}. Starting from the SVD approach illustrated in Fig. \ref{fig:mode_trajectory_protocol}, a linear DMD operator $\mathbf{A}$ is regressed using data from the first $r$ rows of the time-series projection matrix $\mathbf{V}$:
\begin{equation}
\begin{split}
\dot{\mathbf{V}} = \mathbf{A}\mathbf{V}
\end{split}
\end{equation}
The resultant model can reproduce the dynamics in the space of the SVD modes by integrating this differential equation (or using the analytic solution from Eq. \ref{eq:dmd_sol}). It also enables state forecasting simply by integrating further out to some future time. Results can be transformed back into the original delay-coordinate space via matrix multiplication: $\hat{\mathbf{H}} = \mathbf{U}\mathbf{S}\hat{\mathbf{V}}^T$, where $\hat{\mathbf{V}}$ is the DMD reconstruction of the data matrix $\mathbf{V}$.

The Hankel matrix formed by the embedding procedure of Eq.~(\ref{eq:delay_embed}) is guaranteed by construction to have redundant values along matrix diagonals $\{H_{ij}$ where $i+j = \rm(const)\}$. No such property is assured for the DMD reconstruction $\hat{\mathbf{H}}$, so the task of collapsing the resultant signal back down into the original state space is nontrivial. There is an approach used in SSA known as Hankelization, or diagonal averaging, in which a state-space reconstruction $\hat{\mathbf{x}}$ is obtained from $\hat{\mathbf{H}}$ by averaging over the elements which would in a true Hankel matrix be redundant \cite{hassani07}. However, in the case of model forecast residuals discussed in the following section, we find there is often coherent oscillatory behavior along these diagonals which can be obfuscated by averaging. In these cases we therefore simply sample the first $n$ rows of $\hat{\mathbf{H}}$ to compress results back into the state space. We find that the delay registers largely differ from one another only by a global phase factor, so the choice to use the first register is not particularly consequential. 

\section{Data-Driven Decomposition of Nonlinear Systems into Forced Linear Models}
DMD generates models with complex eigenvalues, but for real and stationary input data its modes converge to pure-imaginary conjugate pairs. In this case, a rank-$r$ DMD model yields dynamics at at most $r/2$ distinct frequencies. Thus, while Hankel DMD can produce a linear model which reproduces a discrete-spectrum nonlinear oscillator to arbitrary precision, it will unavoidably fall short when it comes to systems with continuous spectra. In this section, we present an extension of Hankel DMD to model nonlinear dynamics as a forced linear system, for which the model and the forcing signal are learned simultaneously. This approach retains the benefits of linearity for the portion of the model which captures the dominant, (quasi-)periodic dynamics, while adding the versatility to simulate a much broader class of nonlinear systems. 

A standard delay-coordinate DMD model is first generated as explained in the previous section. Working in the space of PCT modes, this means learning the linear operator ${\bf A}$ which offers a best-fit solution to $\dot{\mathbf{v}}(t) = \mathbf{A}\mathbf{v}(t)$ over the full duration of available data. The resulting model is then used for stepwise forecasting: given some $\mathbf{v}(t_j)$, the succeeding observation $\mathbf{v}(t_{j+1})$ can be approximated as follows:
\begin{equation}
\begin{split}
\hat{\mathbf{v}}(t_{j+1}) &= \mathbf{v}(t_j) + \Delta t \mathbf{A}\mathbf{v}(t_j)
\label{eq:stepwise_forecast}
\end{split}
\end{equation}
(This simple Euler time step could be replaced by any forward numerical integration method). Iterating over the first $(m-1)$ columns of $\mathbf{V}$, this approach can be used to construct a full single-step prediction matrix $\hat{\mathbf{V}}$:
\begin{equation}
\begin{split}
\hat{\mathbf{V}} &= \left[ \begin{matrix}
	| & | & | & \\
	\hat{\mathbf{v}}(t_1) & \hat{\mathbf{v}}(t_2) & \hat{\mathbf{v}}(t_3) & \cdots\\
	 | & | & | &
\end{matrix}\right]
\end{split}
\end{equation}
The discrepancy between $\mathbf{V}$ and $\hat{\mathbf{V}}$ can be thought of as a quantitative account of the shortcoming of the DMD representation. We assert that the true dynamics can be written in the form
\begin{equation}
\begin{split}
\dot{\mathbf{v}} &= \mathbf{A}\mathbf{v} + \mathbf{u}(t)
\label{eq:control_model}
\end{split}
\end{equation}
for some parametric exogenous forcing $\mathbf{u}(t)$. In this case a forward Euler step would yield
\begin{equation}
\begin{split}
\mathbf{v}(t_{j+1}) &= \mathbf{v}(t_j) + \Delta t \mathbf{A}\mathbf{v}(t_j) +\Delta t \mathbf{u}(t_j)
\end{split}
\end{equation}
and so
\begin{equation}
\begin{split}
\mathbf{v}(t_{j+1}) - \hat{\mathbf{v}}(t_{j+1}) = \Delta t \mathbf{u}(t_j)
\end{split}
\end{equation}
leading to
\begin{equation}
\begin{split}
\mathbf{u}(t) &= \frac{\mathbf{v}(t+\Delta t) - \hat{\mathbf{v}}(t+\Delta t)}{\Delta t} .
\end{split}
\end{equation}

\begin{figure}[t]
\centering
\includegraphics[width=\columnwidth]{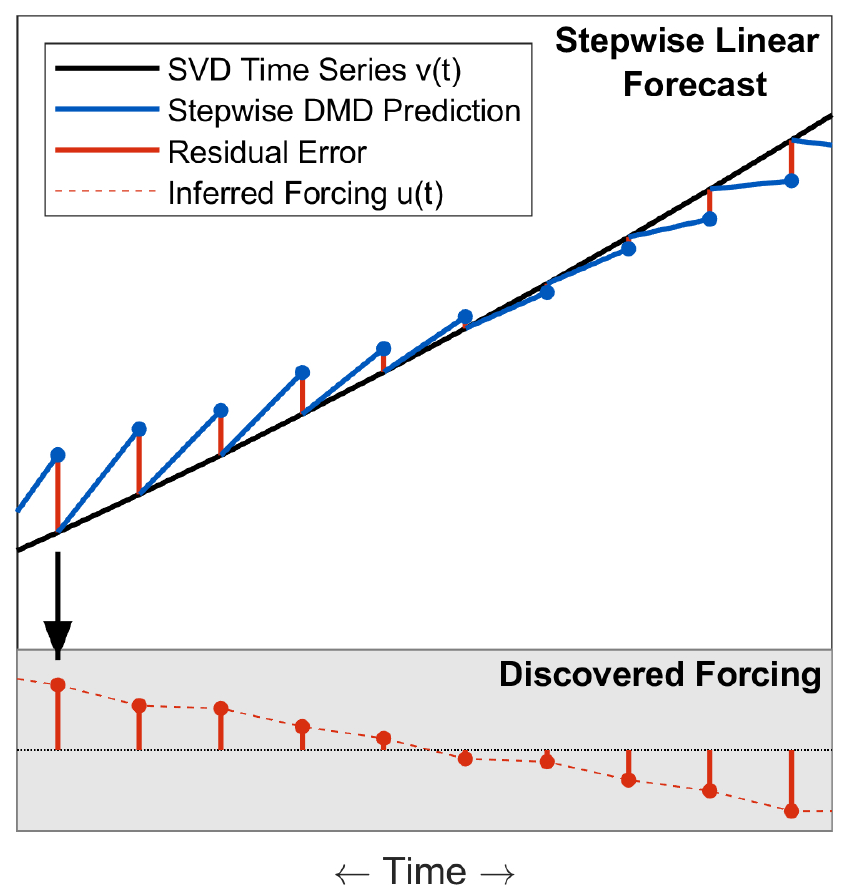}
\caption{Extracting stepwise DMD forecast error to construct a forcing time series. Actuated by this signal, the linear DMD model will perfectly reproduce the true nonlinear dynamics over the observation interval. The model error shown below in red has been scaled up for visual clarity.}
\label{fig:control_signal_protocol}
\end{figure}

An inferred forcing signal can thus be extracted simply by computing the stepwise forecast $\hat{\mathbf{V}}$ using Eq.~(\ref{eq:stepwise_forecast}). This process is illustrated schematically in Fig. \ref{fig:control_signal_protocol}. In one sense this process is tautological: the learned $\mathbf{u}(t)$ takes on whatever value is required to match prediction to reality at a given time. The control model posited in Eq.~(\ref{eq:control_model}) is fully accurate by construction. However, the decomposition it achieves is nontrivial. For any system whose dynamics are dominated by near-periodic evolution (or linear combinations thereof), Hankel DMD can generate a model which captures a large fraction of observed behavior. Under these circumstances, and assuming the rank of ${\bf A}$ is sufficient to account for all prominent spectral peaks, the discovered forcing consists solely of the continuous-spectrum mixing dynamics which lie beyond the reach of DMD. In other words, to the extent that Hankel DMD can be considered an optimal means of discovering a set of $r$ observables in whose space dynamics are most nearly linear (known as a Koopman invariant subspace), ${\bf u}(t)$ offers a window into the residual dynamics which unfold in the orthogonal complement to this subspace.

\begin{figure*}[t]
\centering
\includegraphics[width=\textwidth]{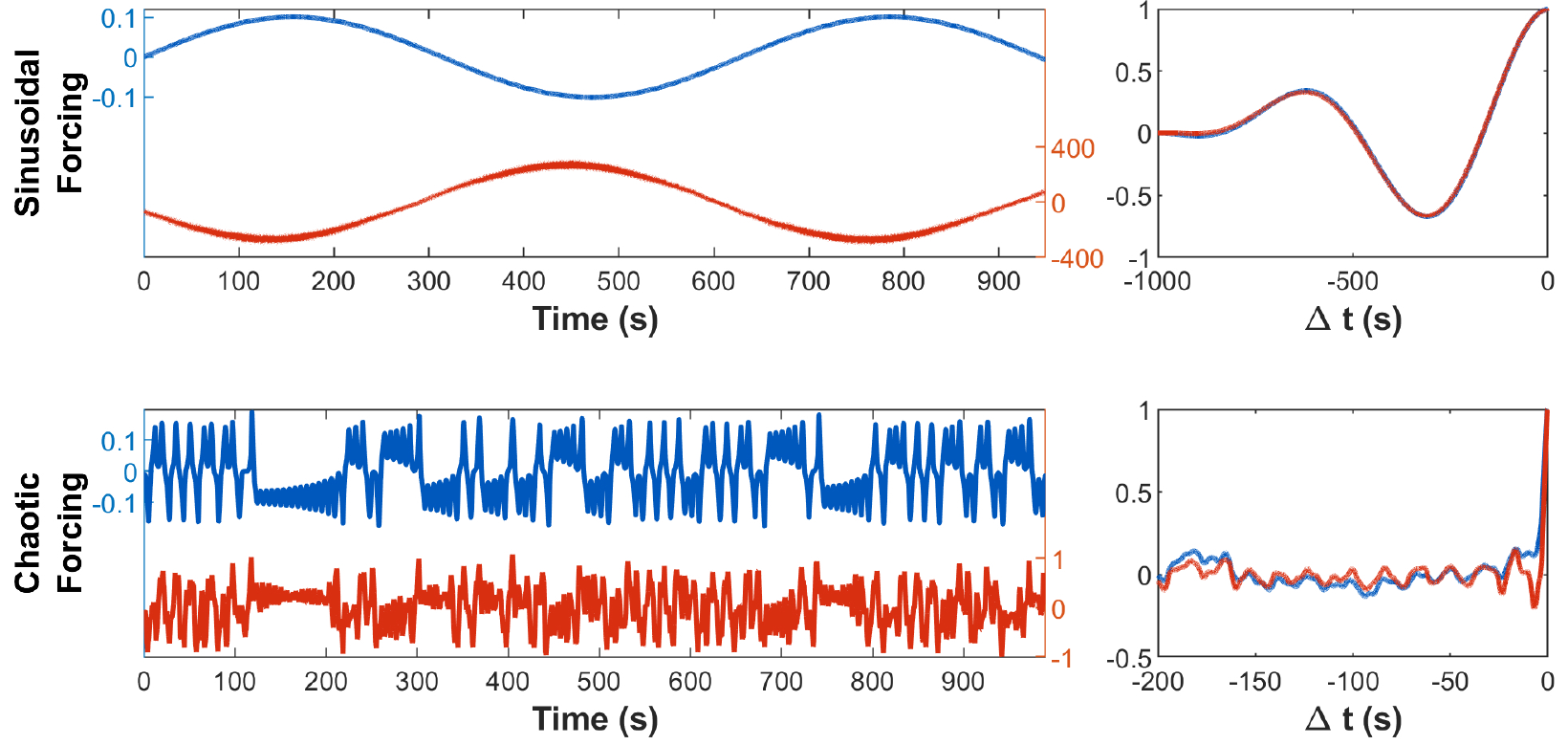}
\caption{Left: discovered forcing signals (red, lower) plotted alongside the true signals (blue, upper) applied to $y_2$ in the simulations. Right: autocorrelation profiles illustrate their similarity up to a global phase factor.}
\label{fig:discovered_forcings}
\end{figure*}

As an example, we consider the forced Van der Pol system defined by:
\begin{equation}
\begin{split}
\dot{y}_1 &= y_2\\
\dot{y}_2 &= \left(1-y_1^2\right)y_2 - y_1 + u(t)
\end{split}
\label{eq:forced_vdp}
\end{equation}
As plotted in Fig. \ref{fig:power_spectrum_comparison}, we consider cases where the forcing $u(t)$ is zero (unforced), where it is sinusoidal, and where it is a chaotic oscillatory signal derived from the $z$ coordinate of a trajectory on the canonical Lorenz attractor. In each case, a Hankel DMD model is generated from data generated over an interval of $1000$ time units sampled at a resolution of $0.01$ time units. The Hankel matrix is constructed using $d = 256$ delay embeddings spaced $q = 4$ time steps apart, for an embedding period of $T^d = qd\Delta t = 10.24$ (which is on the same order as the Van der Pol oscillatory period). The results of the forcing discovery procedure are plotted side by side with the true forcings used to simulate the data in Fig. \ref{fig:discovered_forcings}. While the phases of the corresponding signals clearly do not match, their structures are quite similar and their autocorrelations plotted on the right match very closely. 

The DMD models used to generate Fig. \ref{fig:discovered_forcings} were selected from models with ranks varying from $2$ to $48$. The results plotted (models of $r=12$ and $r=6$, respectively) were chosen as particularly clean representations of the desired results. Some of the models of other ranks produced forcing signals which were somewhat obfuscated by additional high-frequency oscillation. This is because the ability of this method to separate periodic and aperiodic contributions to observed dynamics is limited by the capacity of DMD to fully capture the periodic behavior without any additional, spurious oscillatory modes. The strength of a DMD model in this regard can be evaluated by comparison between its eigenvalue spectrum and the Fourier spectrum of the input data: there should be eigenfrequencies corresponding to all prominent Fourier peaks no more than that. In Fig. \ref{fig:power_spectrum_comparison} it is apparent that the sinusoidal forcing introduced additional peaks at half-integer harmonics of the base frequency, so it is not surprising that a higher-rank DMD model was required to isolate a clean forcing signal relative to the chaotically forced system.

It should be noted that in the case of sinusoidal forcing, the separation that has occurred is not between periodic and aperiodic dynamics: the forcing signal itself is, of course, periodic. But ${\bf u}(t)$ oscillates at a frequency orders of magnitude lower than those which dominate the Fourier power spectrum, and DMD is well known to be ill-equipped to capture dynamics at such disparate time scales \cite{dylewsky2019}. The obtained linear model therefore ignores that slow oscillatory behavior and fully relegates it to the exogenous forcing.

\section{Integration with DMDc: Discovering Linear Control Systems Without Prior Knowledge of the Forcing Signal}
\label{sec:dmdc}

DMD with control (DMDc) extends traditional DMD to fit a model of the form~\cite{proctorbruntonkutz2016}:
\begin{equation}
\begin{split}
\dot{\mathbf{x}} &= \mathbf{A}\mathbf{x} + \mathbf{B}\mathbf{u}(t).
\end{split}
\end{equation}
The algorithm offers a means of simultaneous regression of the intrinsic linear dynamics $\mathbf{A}$ and the control mapping $\mathbf{B}$, but requires a known control signal $\mathbf{u}(t)$ in addition to the state data $\mathbf{x}(t)$. In many circumstances this information is not available, either because a known forcing variable cannot be measured or because the exogenous influence on the system is carried by many environmental variables which cannot even be enumerated, much less measured.	
The control architecture can be extended to a Koopman theoretic framework as well~\cite{KordaMezic2018,Peitz2017arxiv,Kaiser2017arxiv,proctor2018generalizing}.

Incorporating our method for unsupervised forcing discovery into the DMDc approach allows for the construction of linear control models from time series measurements of the state alone. This is accomplished by first running the standard DMD algorithm and collecting the stepwise forecast errors as explained in the previous section, and then performing a second decomposition this time using the DMDc formalism to obtain a modified $\mathbf{A}$ approximating the intrinsic linear dynamics and a control mapping $\mathbf{B}$. The latter is not strictly necessary; the method by which the control signal is generated implicitly assumes equally-weighted forcing on all state variables, so working in the space of the truncated SVD on delay coordinates it can be assumed that $\mathbf{B}$ is the identity ${\bf I}_r$. There exists a variation on DMDc which assumes $\mathbf{B}$ is known, but even the general algorithm reliably discovers this to a good approximation, so there is little practical difference.

\begin{figure*}[t]
\centering
\includegraphics[width=\textwidth]{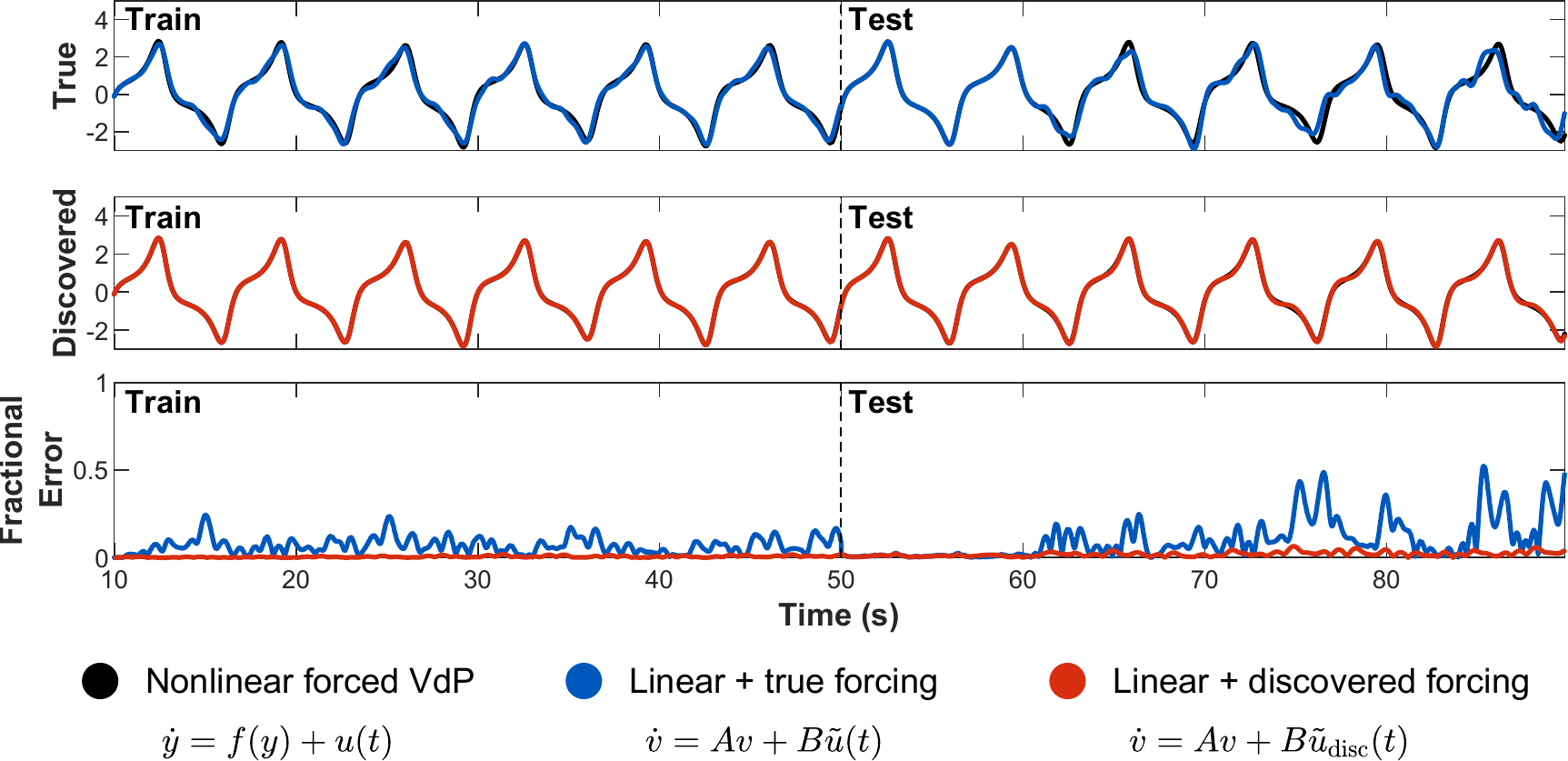}
\caption{$u(t)$ is the control signal (in state space) used to generate the full nonlinear signal (black, underlaid). $\tilde{u}$ is the same control signal lifted into delay embedding space and projected onto the top $r$ SVD modes of the state Hankel matrix ${\bf H}$. $\tilde{u}_{\rm{disc}}$ is the discovered forcing obtained by the procedure diagrammed in Fig. \ref{fig:control_signal_protocol}.}
\label{fig:DMDc_comparison}
\end{figure*}

Figure~\ref{fig:DMDc_comparison} shows results from the application of this method to the chaotically forced Van der Pol system used in the previous section. Two DMDc models are trained: one using the ground truth forcing used in the initial simulation, and one using the discovered forcing obtained from the delay DMD method. In order to apply the original forcing to the model in rank-reduced SVD delay coordinates, it is transformed $\mathbf{u}(t) \in \mathbb{R}^2 \rightarrow \tilde{\mathbf{u}}(t) \in \mathbb{R}^r$ by time-delay embedding and projecting the result onto the top $r$ PCT modes. 

Note that while the DMDc models are trained on the first half of the plotted time series and tested on the remainder, these results should not be interpreted as true forecasts. The models are supplied with forcing data for the full duration, derived either from the original simulation (true forcing) or from the first pass of DMD without control (discovered forcing). This effectively grants the models some foreknowledge of impending dynamics which would not be available for future-state prediction tasks. Rather, the plots of Fig. \ref{fig:DMDc_comparison} are intended to exhibit the results of the DMDc method applied to a system for which the dominant contribution to the stepwise forecast residual is a known external actuation. The plots illustrate the method’s ability to reconstruct observed dynamics equally well with or without ground truth knowledge of the underlying actuation. 

\section{Application: Discovering forcing on real-world data with inherent periodicity}
\label{sec:power_grid}
As an example of the practical diagnostic utility of this method, we apply it to power grid demand data published as part of the RE-Europe data set \cite{jensen15}. Electrical load is sampled hourly over a period from 2012 through 2014. Data is presented for many geolocated nodes of a reduced network representation of the European grid, which we average over to obtain a single time series for each country surveyed. 

This subject was chosen as an example of a real-world data series which contains strong periodic signatures: daily, weekly, and yearly cycles are all quite obviously present, and indeed dominate much of the observed behavior. But the dynamics are of course not entirely periodic; electricity demand is mediated by collective human behaviors, which are in turn influenced by weather patterns, economic trends, and uncountably many other variables. Casting these dynamics as a linear control system in time-delay coordinates effectively separates these factors, with predictable periodicities represented in eigenvalues of the DMD $\mathbf{A}$ matrix and everything else contained in the discovered control signal $\tilde{\mathbf{u}}(t)$. There is no ground truth against which to validate results, as the ``true'' forcing is not a measurable quantity, but we can offer some qualitative observations which corroborate the idea that $\tilde{\mathbf{u}}(t)$ should encode anomalies in human behaviors as they pertain to electricity consumption.

\begin{figure*}[t]
\centering
\includegraphics[width=\textwidth]{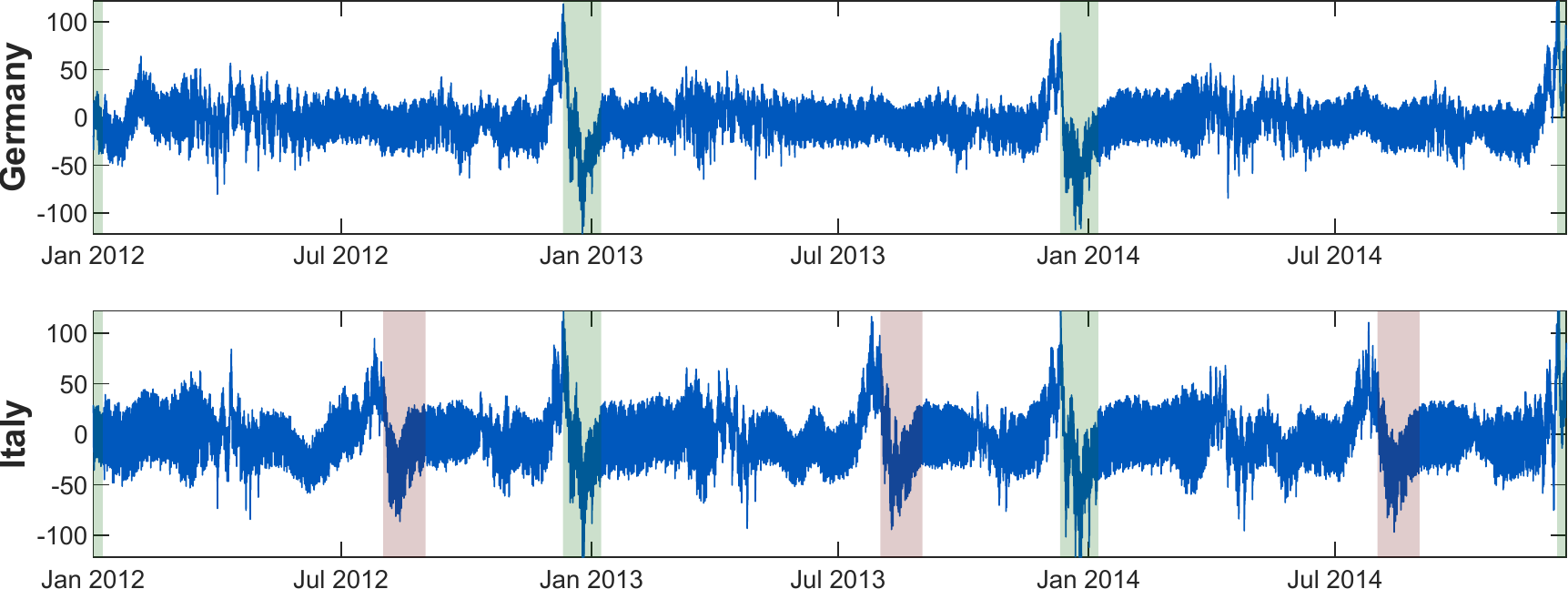}
\caption{Discovered forcing signal (in MW) for mean grid demand per network node in Germany (top) and Italy (bottom). The most obvious feature of the forcing signal is a large annual spike around Christmas in both countries, and an additional annual spike in late summer in Italy. Many businesses in Italy close in August. For reference, the weeks around Christmas (Dec. 11 - Jan. 8) and the month of August (Aug. 1 - Sept. 1) are shaded.}
\label{fig:power_grid_plot}
\end{figure*}

Results are plotted in Fig. \ref{fig:power_grid_plot} for the load data from Germany and Italy. As discussed previously, DMD (like most numerical methods) cannot easily accommodate models with dynamics on highly disparate timescales, such as daily and yearly oscillations. To circumvent this issue we run decompositions on a sliding window of approximately 2 weeks over the full sampling period. This is a sufficiently narrow domain to render seasonal variations negligible over the course of any given sub-series. Forcing signals are then inferred individually for each windowed iteration and then averaged to form a global $\tilde{\mathbf{u}}(t)$ for the full two-year span. 

The most salient features in the results are large annual spikes in which forcing takes on an anomalously high value followed immediately by an anomalous low. Both nations exhibit this pattern once per year in December, and Italy additionally evinces a similar phenomenon (albeit slightly horizontally stretched) in the late summer of every year. This is highly consistent with known behavioral signatures in these countries: both celebrate Christmas with school holidays, business closures, etc., and August is widely observed as a national vacation month in Italy. That these breaks in human routine lead to by far the most prominent spikes in the learned forcing signals (while their influence is overshadowed by standard daily oscillations in the original data) suggests even in the absence of ground truth that this method has to some nontrivial extent successfully disambiguated the quasilinear oscillations which dominate the load data from anomalous activity resulting from more complex environmental variables. This could have far-reaching implications for forecasting of time series such as these, in that it separates the more easily predicted, relatively deterministic dynamics from aperiodic factors which may be better modeled stochastically.

The success of this method in recovering the $u(t)$ signal of the Van der Pol system in Eq. \ref{eq:forced_vdp} was owed in part to the contrived simplicity of this example. For more generic input data, the discovered forcing is not solely determined by true exogeneous actuation of the underlying system. The learned signal also incorporates contributions from any endogenous chaotic (continuous-spectrum) dynamics which cannot be reproduced by the linear Koopman model, as well as any sampling errors that might originate from an imperfect DMD model regression or measurement noise. The Van der Pol example from Fig. \ref{fig:discovered_forcings} minimized these confounding effects because it had a discrete spectrum and the DMD model was regressed on ample, noise-free data. Applying the same procedure to the more complex grid load data demands a slightly more careful interpretation of the results. In a case like this, the difference between external forcing and internal chaotic nonlinearity is not particularly clear. The underlying dynamics governing demand for electricity are complex to the point that there is no obvious distinction between what would qualify as an internal state variable versus an external environmental variable. As a result, it is in many cases philosophically appropriate that the method proposed in this paper makes no attempt to disambiguate between the two. The additional contributions of sampling error and measurement noise also bear acknowledgment, but with enough data to capture all major periodicities of the data the former is likely to be negligible, and the latter would simply pass through and produce a noise floor in the learned forcing (provided that measurement errors are uncorrelated). The spikes in the discovered forcing in the electrical grid load data should therefore be interpreted as affirmative indications of strong disturbances to otherwise nearly-linear dynamics, with the caveat that these disturbances may be endogeneous or exogeneous in origin. It is possible that some further distinction could be made using statistical methods designed to chaotic dynamics from stochastic processes, as in ~\cite{rosso2007} or ~\cite{ravetti2014}, but that is beyond the scope of this paper.

\section{Conclusions}

Since the seminal contributions of Takens, time-delay embeddings of dynamical systems have long been known to contain critical and representative information about the underlying dynamical system measured. In recent years, the advent of high-quality time-series measurements from spatio-temporal systems have afforded the community unprecedented opportunities for constructing data-driven models from such measurement data alone. Not only can time-delay embeddings extract meaningful information from unmeasured latent variables, but such embeddings can be used as a data-driven coordinate system approximating the Koopman operator. The work here advocates the use of these time-delay embeddings for constructing {\em principal component trajectories} (PCT) which provide a time-delay coordinate system which can be used to reconstruct dynamical trajectories via superposition. Indeed, the PCT provide an ideal representation of many dynamical systems and their time-series, especially in systems where an unmeasured latent space is critical to the dynamics. The method is shown to be intimately connected to SSA where various theoretical guarantees are available for infinite time-delay embeddings.

PCT also provide a coordinate system that can be used in conjunction with the DMD algorithm for producing good Koopman operator approximations. It can even be augmented to include external forcing terms in order to model the effects of a continuous spectrum. Such forcing terms are determined and constructed in a completely unsupervised fashion, allowing for a data-driven discovery process that can produce Koopman models which approximate both discrete and continuous spectral dynamics. We offer a demonstration of this method applied to real-world power grid load data to show its utility for diagnostics and interpretation on systems in which somewhat periodic behavior is strongly actuated by unknown and unmeasurable environmental variables. This example illustrates how a completely data-driven method and PCT coordinates can be used in practice. 
It will be interesting to apply this approach to analyze other complex systems, such as for fluid flow control~\cite{Brunton2015amr}, which would benefit from the disambiguation of the linear dynamics and external forcing. Extending these approaches to more complex systems is an area of future research.

\begin{acknowledgments}
We acknowledge support from the UW Engineering Data Science Institute, NSF HDR award 1934292.
SLB further acknowledges support from the Army Research Office (ARO W911NF-17-1-0306). 
JNK further acknowledges support from the Air Force Office of Scientific Research (AFOSR) grant FA9550-17-1-0329.   
\end{acknowledgments}


\bibliography{time_delay_dmd_paper_v4}

\end{document}